\documentstyle[aps,prb]{revtex}

\def\scatt{{\text{scatt}}}		
\def\mono{{\text{mono}}}		
\def\monoalpha{{\text{mono-}\alpha}}	
\def\cut{{\text{cut}}}			

\begin{document}
\draft
\title{A scaling approximation for structure factors in the integral equation
theory of polydisperse nonionic colloidal fluids.}
\author{Domenico Gazzillo, Achille Giacometti}
\address{INFM Unit\`a di Venezia and Facolt\`a di Scienze,
Universit\`a di Venezia,\\
Calle Larga S. Marta DD 2137, I-30123 Venezia, Italy}
\author{Raffaele Guido Della Valle, Elisabetta Venuti}
\address{Dipartimento di Chimica Fisica e Inorganica, Universit\`a di Bologna,\\
Viale Risorgimento 4, I-40136 Bologna, Italy}
\author{Flavio Carsughi}
\address{INFM Unit\`a di Ancona and Facolt\`a di Agraria,\\
Universit\`a di Ancona, Via Brecce Bianche, I-60131 Ancona, Italy}
\date{\today}

\maketitle

\begin{abstract}
Integral equation theory of pure liquids, combined with a new ``scaling
approximation'' based on a corresponding states treatment of pair correlation
functions, is used to evaluate approximate structure factors for colloidal
fluids constituted of uncharged particles with polydispersity in size and
energy parameters. Both hard sphere and Lennard-Jones interactions are
considered. For polydisperse hard spheres, the scaling approximation is
compared to theories utilized by small angle scattering experimentalists
(decoupling approximation and local monodisperse approximation) and to the
van der Waals one-fluid theory. The results are tested against predictions
from analytical expressions, exact within the Percus-Yevick
approximation. For polydisperse Lennard-Jones particles, the scaling
approximation, combined with a ``modified hypernetted chain'' integral
equation, is tested against molecular dynamics data generated for the present
work. Despite its simplicity, the scaling approximation exhibits a
satisfactory performance for both potentials, and represents a considerable
improvement over the above mentioned theories. Shortcomings of the proposed
theory, its applicability to the analysis of experimental scattering data,
and its possible extensions to different potentials are finally discussed.
\end{abstract}

\section{Introduction}

Colloidal solutions consist of mesoscopic particles, with diameters between
$10^{-8}$ m and $10^{-4}$ m, suspended in a microscopic fluid. Colloidal
particles of a same chemical species are not necessarily identical, but may
differ in size, charge or other properties.\cite{Pusey1991,Nagele1996} This
phenomenon is called ``polydispersity'', and systems in which it does not
occur are said to be ``monodisperse''. A microscopic description of a fluid
with a significant polydispersity is a difficult task, requiring a large
number of independent variables. A great simplification is, however,
possible by using the so-called {\it continuous-mixture}, or {\it
polydisperse-mixture}, formalism. Such a model, adopted also in this paper,
views the fluid as containing an infinite number of components
($p\rightarrow\infty$), with a continuous distribution of size and/or other
properties. All molar fractions of the components are then replaced by a
single distribution function, which describes the composition of the
system.

Since polydispersity can significantly affect the microscopic ordering of
colloidal suspensions, it must be taken into account in the analysis of
experimental data on such fluids. In particular, we are interested in static
structure factors obtainable from small angle scattering of light, neutrons
or X-rays.

In this paper we present integral equation (IE) calculations for the
structure of polydisperse one-species fluids constituted of uncharged
particles with hard sphere (HS) or Lennard-Jones (LJ) interactions, with
polydispersity in size and, for LJ systems, also in energy parameters. IEs of
statistical mechanics represent a powerful, although approximate, tool to
determine both structures and thermodynamics of fluids in a simple
way. However, while using IEs for pure fluids or binary mixtures is a rather
common and successful practice, their application to multicomponent systems
with large $p$ or $p\rightarrow\infty$ is problematic and, consequently, less
frequent in the literature.\cite{Gazzillo1994,Gazzillo1995} Only in special
cases, when analytical solutions are available, IE calculations for
polydisperse fluids can be performed in a rigorous and relatively easy
way. For instance, a closed analytical formula for the scattering intensity of
polydisperse hard spheres was obtained by Vrij \cite{Vrij1978} from the solution
of the Percus-Yevick (PY) IE. Similar expressions were derived for
polydisperse charged hard spheres, by using the corresponding analytical
solution in the mean spherical approximation.\cite{Gazzillo1997}

Unfortunately, for most ``closures'' and for most potentials, including the
LJ one, the IEs must be solved numerically, requiring more and more computer
memory and time with increasing $p$, so that the problem soon becomes
practically intractable (D'Aguanno {\it et al.} \cite{Aguanno1991} reached
$p=10$ for Yukawa plus HS interactions).

To overcome the impossibility of investigating polydisperse systems when IEs
have to be solved numerically, alternative routes have been proposed. The
common idea is to replace the polydisperse fluid with an appropriate system
with very few components. A first method, not analyzed in this paper, builds
up an equivalent {\it effective mixture} with $p^\prime\ll p$ new components,
whose molar fractions and diameters - in the case of size polydispersity - are
determined by replacing the continuous size distribution with a
$p^\prime$-component histogram and requiring the equality of the first
$2p^\prime$ moments of the two distributions ($p^\prime=2$ or 3 is usually
sufficient).\cite{Nagele1996,Aguanno1992} Other methods assume that the
properties of polydisperse fluids can be obtained, to a good approximation,
from those of {\it pure} fluids, if suitably averaged parameters and
appropriate recipes are used. This same basic idea has often been employed, in
several variants (van der Waals equation of state, perturbation theories,
etc.), to predict thermodynamics of liquid mixtures with a small number of
components.\cite{Smith1973,Mcdonald1973,Lee1988} Some significant applications of
these concepts have also been made to structural studies of polydisperse
systems, mainly by experimentalists in the analysis of scattering
data. Kotlarchyk and Chen \cite{Kotlarchyk1983} introduced the {\it decoupling
approximation} (DA), which is perhaps the most famous of these approaches and
involves an exact evaluation of the form factors for all components of the
polydisperse system, but approximates every partial structure factor with that
of an one-component fluid. A second approximation, suggested by Pedersen \cite{Pedersen1994} and known as {\it local monodisperse approximation} (LMA),
replaces the polydisperse fluid not with a single pure system, but with a
superposition of non-interacting pure fluids, whose number equals that of the
species in the mixture.

The aim of the present paper is twofold. First, both DA and LMA are
discussed in terms of pair correlation functions, to get some insight into
their shortcomings. Second, we propose a simple {\it scaling approximation}
(SA), which, with respect to DA and LMA, takes excluded volume effects more
correctly into account and therefore yields significantly improved
structure factors. By using corresponding states arguments the SA derives
all pair correlation functions of a polydisperse mixture from an
appropriate pure fluid counterpart, at the cost of {\it only one} IE
computation. The performance of SA is tested on two typical potential
models, namely polydisperse hard spheres and polydisperse Lennard-Jones
particles. For HS systems, SA results for the ``measurable'' structure
factor are compared to those obtained from Vrij's analytical
expression,\cite{Vrij1978} which is exact within the PY approximation. For
polydisperse LJ fluids, SA is tested against molecular dynamics data
generated for this purpose.

\section{Integral equation theory}

\subsection{Basic equations}

The Ornstein-Zernike (OZ) integral equations of the liquid state theory for
$p$-component mixtures with spherically symmetric interparticle potentials
are \cite{Hansen1986}

\begin{equation}
h_{\alpha\beta}\left(r\right) = c_{\alpha\beta}\left(r\right)
+ \rho \sum_{\gamma=1}^p x_\gamma \int {\rm d}{\bf r}^\prime ~
  c_{\alpha\gamma}\left(r^\prime \right) ~
  h_{\gamma\beta}\left(|{\bf r-r}^\prime |\right),
\end{equation}

\noindent
where $h_{\alpha\beta}\left(r\right) \equiv g_{\alpha\beta}\left(r\right)-1$
is the total correlation function between two particles of species $\alpha$
and $\beta$ at a distance $r$,
$g_{\alpha\beta}\left(r\right)$ is the radial distribution function (RDF),
$c_{\alpha\beta}\left(r\right)$ is the direct correlation function, $\rho
\equiv N/V$ the total number density ($N$ = total particle number, $V$ =
volume) and $x_\gamma$ the molar fraction of species $\gamma$. These
equations can be solved only when coupled with a {\it closure} relationship,
given by the exact formula

\begin{equation}
c_{\alpha\beta}\left(r\right) =\exp \left[ -u_{\alpha\beta}\left(r\right)
/k_BT+\gamma_{\alpha\beta}\left(r\right) +B_{\alpha\beta }\left(r\right)
\right] -1-\gamma_{\alpha\beta}\left(r\right),
\end{equation}

\noindent
plus an approximation to the ``bridge'' functions
$B_{\alpha\beta}\left(r\right)$, which are functionals of
$h_{\alpha\beta}\left(r\right)$ and higher order correlation functions \cite{Hansen1986,Gazzillo1993} ($u_{\alpha\beta}\left(r\right)$ is the
interparticle potential, $k_B$ is Boltzmann's constant and $T$ the absolute
temperature; $\gamma_{\alpha\beta}\left(r\right) \equiv h_{\alpha\beta}
\left( r\right) -c_{\alpha\beta}\left(r\right)$). The OZ equations admit an
analytical solution only in a relatively small class of cases, for some
potentials and some peculiar closures.

In the first case considered in this paper, i.e., for hard sphere (HS)
particles with additive diameters \cite{Lebowitz1964} $\sigma_\alpha$,
corresponding to the potential

\begin{equation}
u_{\alpha\beta}\left(r\right) =\left\{
\begin{array}{lll}
+\infty, & & r<\sigma_{\alpha\beta} \equiv (\sigma_\alpha +\sigma_\beta)/2 \\
0, & & r\geq \sigma_{\alpha\beta}
\end{array}
\right. \label{hs}
\end{equation}

\noindent
an analytical solution is possible if one adds to the exact hard core
condition, $h_{\alpha\beta}\left(r\right)=-1$ for $r<\sigma_{\alpha\beta}$,
the Percus-Yevick (PY) approximation

\begin{equation}
B_{\alpha\beta}\left(r\right) =
\ln \left[ 1+\gamma_{\alpha\beta}\left(r\right) \right] -
\gamma_{\alpha\beta}\left(r\right) \quad \text{for~~} r>\sigma_{\alpha\beta},
\end{equation}

\noindent
which is equivalent to $c_{\alpha\beta}\left(r\right)=0$ for
$r>\sigma_{\alpha\beta}$.

Only a numerical solution is feasible in the second case
of this paper, i.e., the Lennard-Jones (LJ) potential,

\begin{equation}
u_{\alpha\beta}\left(r\right) =4 ~ \varepsilon_{\alpha\beta}\left[
\left(\frac{\sigma_{\alpha\beta}}r\right)^{12} -
\left(\frac{\sigma_{\alpha\beta}}r\right)^6\right],
\end{equation}

\noindent
where $\sigma_{\alpha\beta}$ are LJ diameters and
$\varepsilon_{\alpha\beta}$ energy parameters (well depths). The number of
independent LJ parameters is reduced by assigning individual parameters
$\left(\sigma_\alpha,\varepsilon_\alpha \right)$ to each species $\alpha$
and obtaining the cross-interactions from combination rules

\begin{eqnarray}
\sigma_{\alpha\beta} \equiv & (\sigma_\alpha+\sigma_\beta)/2
  \qquad & \text{(Lorentz rule)},   \label{pl1} \\
\varepsilon_{\alpha\beta} \equiv & \sqrt{\varepsilon_\alpha \varepsilon_\beta}
  \qquad & \text{(Berthelot rule)}. \label{pl2}
\end{eqnarray}

\noindent
The OZ equation is solved using the {\it modified hypernetted chain} (MHNC)
closure described in Ref. \onlinecite{Gazzillo1993}, which is one of the best IE
approximations for both structures and thermodynamics of LJ one-component
fluids. All versions of the MHNC theory assume that the bridge function $B(r)$
has roughly the same functional form for all potentials and replace the
unknown $B(r)$ of the system under study with that of a reference system whose
properties are known. We approximate the LJ one-component bridge function with
that of an appropriate HS fluid, i.e., $B_{LJ}(r;\sigma,\rho,T)\simeq
B_{HS}(r;d,\rho)$, where $d $ is an equivalent HS diameter, which depends on
$\rho$ and $T$. Our choice for $B_{HS}$ is a slight modification of an
empirical analytical approximation proposed by Malijevsky and Labik \cite{Malijevsky1987,Labik1989} (ML), and $d$ is selected by equating the second
density derivative of the free energy of the LJ fluid with that of the HS
reference, according to a prescription due to Rosenfeld and
Blum.\cite{Rosenfeld1986} Further details can be found in the original
paper.\cite{Gazzillo1993}

\subsection{Structure factors}

\noindent
The Ashcroft-Langreth \cite{Ashcroft1967} {\it partial} structure factors
$S_{\alpha\beta}(q)$ are defined as

\begin{equation}
S_{\alpha\beta}(q) = \delta_{\alpha\beta}
  + \rho ~ \sqrt{x_\alpha x_\beta} ~ \widetilde{h}_{\alpha\beta}(q),
\end{equation}

\noindent
where $\delta_{\alpha\beta}$ is the Kronecker delta and
$\widetilde{h}_{\alpha\beta}(q)$ the three-dimensional Fourier transform of
$h_{\alpha\beta}\left(r\right)$. Appropriate linear combinations of
partial structure factors define {\it global} structure factors. A first
example of these is the ``measurable'' structure factor \cite{Nagele1996}

\begin{equation}
S_M(q)=\sum_{\alpha=1}^p\sum_{\beta=1}^p
w_\alpha(q) w_\beta(q)~ \sqrt{x_\alpha x_\beta} ~ S_{\alpha\beta}(q), \label{sm}
\end{equation}

\noindent
with

\begin{equation}
w_\nu(q)\equiv\frac{F_\nu(q)}{\sqrt{\left\langle F^2(q)\right\rangle}},
\label{w}
\end{equation}

\noindent
where $F_\nu(q)$ is the scattering form factor of species $\nu$, and
angular brackets, $\left\langle \cdots \right\rangle$, denote, here and in
the following, compositional averages over the distribution of particles,
i.e., $\left\langle Y\right\rangle\equiv\sum_{\alpha=1}^p x_\alpha Y_\alpha$
for any property $Y$ (note that $\left\langle w^2(q)\right\rangle=1)$.

We assume that the scattering matter has a well-defined boundary, i.e., there
is a {\it scattering core} with a well-defined {\it scattering volume}, not
necessarily coincident with the particle volume. The former, in fact, depends
on the particle-radiation interaction, whereas the latter is determined by
the interparticle repulsions and may even not be well-defined, as for LJ
particles (only molecules with {\it hard} body repulsions possess a
well-defined volume). For spherical homogeneous scattering cores, the form
factors are

\begin{equation}
F_\alpha(q;\sigma_\alpha^\scatt)\propto
V_\alpha^\scatt ~ \frac{3j_1\left(q\sigma_\alpha^\scatt/2\right)}
{q\sigma_\alpha^\scatt/2}, \label{scatt}
\end{equation}

\noindent
where $\sigma_\alpha^\scatt\leq \sigma_\alpha$ is the diameter of the
scattering core of species $\alpha$, $V_\alpha^\scatt=\frac{\pi}{6}\left(\sigma_\alpha^\scatt\right)^3$ its volume, and $j_1(x)=(\sin x-x\cos
x)/x^2$ is the first-order spherical Bessel function. The notation
$F_\alpha(q;\sigma_\alpha^\scatt)$ emphasizes the dependence of the form
factor on $\sigma_\alpha^\scatt$, which in general may differ from
$\sigma_\alpha$. In this paper, for the sake of simplicity,
$\sigma_\alpha^\scatt$ is taken coincident with $\sigma_\alpha$, for all
species and for both HS and LJ potentials.

A second global structure factor of interest is the Bhatia-Thornton
number-number structure factor,\cite{Bhatia1970} obtainable by taking all
$w_\nu=1$ in Eq. (\ref{sm})

\begin{equation}
S_{NN}(q)=\sum_{\alpha=1}^p\sum_{\beta=1}^p \sqrt{x_\alpha x_\beta} ~
S_{\alpha\beta}(q).
\end{equation}

\noindent
While $S_M(q)$ represents the structure factor measured in small
angle scattering experiments, $S_{NN}(q)$ is related to the fluctuations in
particle numbers.

\subsection{Polydisperse continuous limit}

All previous formulas, written in a discrete form, refer to a finite number
$p$ of components. On the other hand, theoretical treatments of
polydispersity with continuous distributions refer to systems with an
infinite number of components ($p\rightarrow \infty)$.

For HS particles polydispersity of only one property - the diameter $\sigma$
- is possible. For LJ particles both $\sigma$ and $\varepsilon$ might be
polydisperse. In this paper, however, to simplify the LJ model, energy and
size parameters will be correlated according to the law \cite{Kofke1986,Kofke1987}

\begin{equation}
\varepsilon_\alpha =\varepsilon_{\left\langle \sigma \right\rangle}\left(
\frac{\sigma_\alpha} {\left\langle \sigma \right\rangle} \right)^z,
\label{pcl13}
\end{equation}

\noindent
where $\left\langle \sigma \right\rangle$ is the average diameter,
$\varepsilon_{\left\langle \sigma \right\rangle}$ the corresponding well
depth, and $z$ is an adjustable exponent, for which we take the value $z=2$.
Other choices will be discussed later.

The polydisperse continuous limit of the previous discrete expressions
can therefore be obtained by simple {\it replacement rules}:

\begin{equation}
x_\alpha \rightarrow {\rm d}x = f(\sigma) {\rm d}\sigma, \label{pcl1}
\end{equation}
\begin{equation}
\sum_\alpha x_\alpha ...\rightarrow \int {\rm d}\sigma f(\sigma) ..., \label{pcl2}
\end{equation}

\noindent
where $f(\sigma){\rm d}\sigma$ is the probability of finding a
particle with diameter in the range $\sigma \div \sigma +{\rm d}\sigma$, and
the distribution function $f(\sigma)$ ({\it molar fraction density
function}) is normalized. Specifically, we shall use the Schulz (or gamma)
distribution \cite{Beurten1981}

\begin{equation}
f(\sigma)=\frac{b^a}{\Gamma(a)}\sigma^{a-1}e^{-b\sigma} \;\;(a>1),
\end{equation}

\noindent
where $\Gamma$ is the gamma function,\cite{Abramowitz1972} and the two parameters
$a$ and $b$ can be expressed as $a=1/s^2$ and
$b=a/\left\langle\sigma\right\rangle$, in terms of the mean value
$\left\langle \sigma \right\rangle$ and the relative standard deviation
$s\equiv \sqrt{\left\langle \sigma^2\right\rangle -\left\langle \sigma
\right\rangle^2}/\left\langle \sigma \right\rangle$. The dispersion parameter
$s$ measures the degree of polydispersity, and varies in the range
$0<s<1$. For $s\rightarrow 0$, the Schulz distribution reduces to a Dirac
delta function centered at $\left\langle\sigma\right\rangle$ (monodisperse
limit). For small $s$ values $f(\sigma)$ is very similar to a Gaussian
distribution (without its drawback of unphysically negative diameters). For
$s$ closer to one, $f(\sigma)$ becomes asymmetric, with a long tail at large
diameters.\cite{Beurten1981} The first three moments of the Schulz distribution
are: $\left\langle \sigma \right\rangle,\left\langle \sigma^2\right\rangle
=\left(1+s^2\right) \left\langle \sigma \right\rangle^2$ and $\left\langle
\sigma^3\right\rangle =\left(1+s^2\right) \left(1+2s^2\right) \left\langle
\sigma \right\rangle^3$.

Whenever analytical integration is impossible, numerical integration brings
back to discrete expressions (with large $p$, of order $10^2-10^3$), and
therefore the replacement rule of Eq. (\ref{pcl2}) becomes unnecessary. Thus
no integral is needed in the formulas and the discrete notation is always
employed, implicitly assuming $x_\alpha=f(\sigma_\alpha)\Delta\sigma$,
which is the discrete analogue of Eq. (\ref{pcl1}) ($\Delta\sigma$ is the
grid size in the numerical integration).

\section{Corresponding states and scaling approximation}

To introduce our scaling approximation for global structure factors, a
corresponding states \cite{Rowley1994,Lucas1991} approach will be used.
The correspondence principle applies to systems which have  {\it
conformal} pair potentials, i.e., potentials of the same {\it
shape}. The principle
takes its simplest form when the potential $u_\alpha$ of each species
$\alpha$, in a set of conformal substances, depends on two parameters
only and can be written as

\begin{equation}
u_\alpha(r)=\varepsilon_{\alpha} u^*\left(\frac r{\sigma_\alpha}
\right), \label{cs1}
\end{equation}

\noindent
where $\sigma_\alpha$ and $\varepsilon_{\alpha}$ are a characteristic length
and a characteristic energy, respectively, while $u^*$ is a
dimensionless function of the dimensionless distance $r^*\equiv
r/\sigma$. Such a form of $u_\alpha(r)$ implies that all
properties of a set of conformal fluids can be written in terms of
dimensionless reduced variables, e.g. temperature $T_\alpha^*\equiv
k_BT/\varepsilon_\alpha$, number density $\rho_\alpha^*\equiv \rho
\sigma_\alpha^3$, and pressure $p_\alpha^*\equiv p_\alpha
\sigma_\alpha^3/\varepsilon_\alpha$.

When written in terms of reduced distance $r_\alpha^*\equiv
r/\sigma_\alpha$ and wavevector $q_\alpha^*\equiv q\sigma_\alpha$,
the RDF and structure factor of any pure fluid of species $\alpha$ can be
derived by scaling as

\begin{eqnarray}
g_\alpha(r;\rho,T;\sigma_\alpha,\varepsilon_\alpha) & =
\widehat{g}\left(r_\alpha^*;\rho_\alpha^*,T_\alpha^*\right),
  \label{cs3} \\
S_\alpha(q;\rho,T;\sigma_\alpha,\varepsilon_\alpha) & =
\widehat{S}\left(q_\alpha^*;\rho_\alpha^*,T_\alpha^*\right),
  \label{cs4}
\end{eqnarray}

\noindent
where $\widehat{g}$ and $\widehat{S}$ are functions common to the
entire set of conformal substances. Eqs. (\ref{cs3}) and (\ref{cs4})
indicate that the scaling correspondence applies not only to
thermodynamic variables, but also to ``positions'' in $r-$ and
$q-$space.\cite{Parrinello1974} Both HS and LJ potentials satisfy the
scaling condition given by Eq. (\ref{cs1}) (in the HS case, since
$\varepsilon_\alpha=1$, RDFs and structure factors do not depend on
$T$).

{\it Conformal mixtures} are those in which all pair potentials are
conformal to each other and to that of a pure (monodisperse
one-species) reference fluid, according to the scaling relation

\begin{equation}
u_{\alpha\beta}(r)=\varepsilon_{\alpha\beta} ~
u_\mono^*\left(\lambda_{\alpha\beta} r_\mono^*\right), \label{cs13}
\end{equation}

\noindent
where $r_\mono^*\equiv r/\sigma_\mono$, while $\varepsilon_{\alpha\beta}$ and
$\lambda_{\alpha\beta}$ are parameters characteristic of the pair
$\alpha$,$\beta$. For mixtures conformality of potentials does not imply
conformality of RDFs in the same simple way as for pure fluids. Nevertheless,
corresponding states arguments have sometimes been employed by postulating
approximate conformality relations between mixture and pure
RDFs.\cite{Lee1988,Rowley1994} The same approach is also followed in the present
paper.

We call {\it scaling approximations} (SA) those assuming approximate
conformality of all RDFs of a mixture, according to the relation

\begin{equation}
g_{\alpha\beta}\left(r;\rho,{\bf x},T;\{\sigma_{\gamma\delta}\},
\{\varepsilon_{\gamma\delta}\}\right) \simeq
\widehat{g}_\mono\left(\lambda_{\alpha\beta}
r_\mono^*;\rho_\mono^*,T_\mono^*\right), \label{cs14}
\end{equation}

\noindent
where ${\bf x}$, $\{\sigma_{\gamma\delta}\},
\{\varepsilon_{\gamma\delta}\}$ represent the complete set of molar
fractions and potential parameters,

\begin{equation}
\rho_\mono^*\equiv \rho \sigma_\mono^3,\qquad T_\mono^*\equiv
\frac{k_BT}{\varepsilon_\mono}, \label{cs15}
\end{equation}

\noindent
while $\sigma_\mono$ and $\varepsilon_\mono$ are suitably chosen average
potential parameters. Generalizations with $\sigma_\mono$ and
$\varepsilon_\mono$ replaced by pair-dependent parameters are
possible,\cite{Lee1988} but we will restrict ourselves to the simplest case. The
value of each $g_{\alpha\beta}$ at $r$ is obtained from a single functional
form, appropriate to a pure HS fluid, by evaluating it at a {\it scaled}
pair-dependent distance, $r_{\alpha\beta}^\prime=\lambda_{\alpha\beta}r$, and
at a {\it corresponding} thermodynamic {\it state.}

Since the Fourier transform of $h_\mono(\lambda_{\alpha\beta} r)$ is
$\lambda_{\alpha\beta}^{-3}~\widetilde{h}_\mono(\lambda_{\alpha\beta}^{-1}q)$
and $S_\mono\left(\lambda_{\alpha\beta}^{-1}q\right) =
\widehat{S}_\mono\left(\lambda_{\alpha\beta}^{-1}q_\mono^*\right)$, with

\begin{equation}
q_\mono^*\equiv q\sigma_\mono,
\end{equation}
\noindent
the approximate partial structure factors become

\begin{equation}
S_{\alpha\beta}(q) \simeq \delta_{\alpha\beta} + ~ \sqrt{x_\alpha x_\beta} ~
 \lambda_{\alpha\beta}^{-3}\left[
 \widehat{S}_\mono\left(\lambda_{\alpha\beta}^{-1}q_\mono^*;
 \rho_\mono^*, T_\mono^*\right) -1 \right]. \label{cs17}
\end{equation}

\section{Hard sphere potential}

For clarity, the various approximations examined in this paper are presented
starting from a particular physical system, namely the HS potential. Both DA
and LMA will be slightly reformulated, to point out the underlying
approximations in terms of pair correlation functions
$g_{\alpha\beta}\left(r\right)$.

\subsection{Decoupling approximation}

Kotlarchyk and Chen \cite{Kotlarchyk1983} proposed the {\it decoupling
approximation} in a rather general form, to treat both polydisperse fluids
and systems of non-spherical particles. They performed two basic
approximations:

a) First, they replaced orientation-dependent interparticle potentials with
spherically symmetric ones. Strictly speaking, this is the actual
``decoupling approximation'', which allows one to break the ensemble average
present in the exact expression of the scattering intensity into two
factors, neglecting correlations between particle orientations and
positions. The result is the Fournet-Vrij expression for the scattering
intensity,\cite{Vrij1978} with form factors averaged over particle
orientations. For spherical particles, this first approximation is
unnecessary and hence will not be exploited in the present paper.

b) Second, the partial structure factors $S_{\alpha\beta}(q)$ were
approximated in terms of the structure factor of an appropriate {\it pure}
fluid, as

\begin{equation}
S_{\alpha\beta}(q)
 = \delta_{\alpha\beta} +\rho ~ \sqrt{x_\alpha x_\beta} ~
\widetilde{h}_{\alpha\beta}(q)\simeq \delta_{\alpha\beta}
+ \sqrt{x_\alpha x_\beta} ~ \left[ S_\mono(q)-1\right], \label{da1}
\end{equation}

\noindent
where the definition of the effective pure fluid must be completed
suitably, depending on the particular physical system. In terms of
correlation functions, this approximation is equivalent to assuming that: i)
all the RDF $g_{\alpha\beta}(r)$ have the same dependence on $r$, being
equal to the RDF of the pure fluid, i.e., $g_{\alpha\beta}(r)\simeq
g_\mono(r)$, or\qquad

\begin{equation}
h_{\alpha\beta}(r)\simeq h_\mono(r); \label{da2}
\end{equation}

\noindent
ii) the number density of the pure fluid is equal to the total
number density of the mixture, i.e., $\rho_\mono=\rho$. With Eq. (\ref{da2})
and $S_\mono=1+\rho_\mono~\widetilde{h}_\mono$, this implies Eq. (\ref{da1}).

For polydisperse HS, Kotlarchyk and Chen \cite{Kotlarchyk1983} defined the
diameter of the effective HS fluid by choosing
\begin{equation}
\sigma_\mono=\left\langle \sigma^3\right\rangle^{1/3},
\end{equation}

\noindent
ensuring that the volume fraction (or packing fraction) of the
pure fluid $\eta_\mono\equiv \left(\pi/6\right)
\rho_\mono\sigma_\mono^3$ is equal to the total volume fraction of the
mixture, $\eta\equiv\left(\pi/6\right)\rho\left\langle\sigma^3\right\rangle$.

\noindent
The resulting DA structure factors are

\begin{eqnarray}
S_{\alpha\beta}(q)\quad & \simeq & \delta_{\alpha\beta} +\sqrt{x_\alpha x_\beta
}~\left[ \widehat{S}_\mono(q_\mono^*;\eta)-1\right],     \label{da3} \\
S_M(q)^{DA} ~ & = & 1+\left\langle w(q)\right\rangle^2\left[
\widehat{S}_\mono(q_\mono^*;\eta)-1\right],              \label{da4} \\
S_{NN}(q)^{DA} & = & \widehat{S}_\mono(q_\mono^*;\eta). \label{da5}
\end{eqnarray}

\noindent
On the left hand side of these equations the dependence on the thermodynamic
state and the potential parameters of the mixture has been omitted for
simplicity. Moreover, here and in the following, we simply write
$\eta_\mono=\eta$ instead of $\rho_\mono^*$ in $\widehat{S}_\mono$, since
these quantities are proportional.

Before concluding this subsection, some remarks are appropriate. The first one
is that the approximation expressed by Eq. (\ref{da2}) was already proposed in
the theory of liquid mixtures: it is known as {\it random mixture}, or
{\it random mixing}, approximation,\cite{Smith1973,Mcdonald1973} and sometimes
is also referred to as {\it substitutional model}.\cite{Nagele1996} The
weakness of this approach is evident from its RDF form, Eq. (\ref{da2}),
because it ignores the ordering which takes place in the presence of different
particle sizes. Therefore, the DA is expected to be a very poor approximation
for moderate or even low size polydispersity. A second remark is that, for a
given interparticle potential $u_\mono$, it is possible to choose among
several routes to evaluate $S_\mono$ from the pure fluid OZ integral equation,
by changing the ``closure''. In the luckiest cases an analytical solution may
be available, but a numerical solution is always feasible. For polydisperse
HS, Kotlarchyk and Chen \cite{Kotlarchyk1983} used the PY analytical solution
for the monodisperse fluid, but other more accurate closures (as for instance,
the Ballone-Pastore-Galli-Gazzillo \cite{Ballone1986} (BPGG) or the Rogers-Young \cite{Rogers1984} (RY) approximations) could also be employed. A final remark is
that DA may be regarded as a scaling approximation, obtained from
Eq. (\ref{cs17}) with the choice $\lambda_{\alpha\beta}=1$ and
$\sigma_\mono=\left\langle \sigma^3\right\rangle^{1/3}$.

\subsection{Local Monodisperse Approximation}

This approximation was originally formulated by Pedersen \cite{Pedersen1994}
for polydisperse hard spheres, but can be easily extended to polydisperse
fluids with different potentials. According to the original presentation, in
LMA a $p$-component mixture is approximated by a set of $p$ non-interacting
pure subsystems (a subscript ``$\monoalpha$'' will be used to characterize that
of species $\alpha$), and the scattering intensity is calculated as a
superposition of the scattering intensities from the subsystems, weighted
according to the size distribution of the mixture.

For HS, LMA may be expressed, in terms of partial structure factors, as

\begin{equation}
S_{\alpha\beta}(q)\simeq \left\{
\begin{array}{ll}
0, & \alpha \neq \beta \\
\widehat{S}_\mono(q_\alpha^*;\eta),\qquad & \alpha =\beta,
\end{array}
\right. \label{lma1}
\end{equation}

\noindent
which implies that

\begin{equation}
{S}_M(q)^{LMA}=\sum_{\alpha=1}^p x_\alpha
~w_\alpha^2(q)~\widehat{S}_\mono(q_\alpha^*;\eta),
\end{equation}

\noindent
${S}_{NN}(q)^{LMA}$ is obtained by putting all $w_\alpha(q)=1$. The pure
subsystem of species $\alpha$ consists of hard spheres with diameter
$\sigma_\monoalpha=\sigma_\alpha$ and at a number density
$\rho_\monoalpha=\rho \left\langle \sigma^3\right\rangle/~\sigma_\alpha^3$,
which differs, in general, from the density $\rho_\alpha \equiv x_\alpha \rho$
of that species in the mixture. Such a choice for the $\rho_\monoalpha$ values
of the $p$ subsystems ensures that the volume fraction of each of these pure
fluids, $\eta_\monoalpha \equiv (\pi/6)\rho_\monoalpha \sigma_\alpha^3$,
equals the total volume fraction $\eta$ of the mixture. For $S_\mono$ Pedersen
used the PY analytical expression.\cite{Pedersen1994}

In terms of pair correlation functions, LMA may be written as

\begin{equation}
\rho ~ \sqrt{x_\alpha x_\beta} ~ h_{\alpha\beta}(r) \simeq
\delta_{\alpha\beta} ~ \rho_\monoalpha ~
\widehat{h}_\mono\left(r_\alpha^*;\eta\right), \label{lma5}
\end{equation}

\noindent
which shows that LMA neglects all interactions between particles with
different diameters, i.e., $h_{\alpha\beta}(r)=0$ if
$\alpha\neq\beta$. Pedersen justified this approximation on the ground of a
physical picture, in some sense complementary to DA, which assumes that
particle sizes and positions are completely correlated. This means that
particle size varies slowly with position, so that every particle is
surrounded by particles of the same size and the system looks {\it locally
monodisperse}. On the other hand, LMA may simply be regarded, in a
corresponding states framework, as a conformality assumption not for
$g_{\alpha\beta}$ but for $S_{\alpha\beta}$, done to reduce
the double sum to a single sum in both structure factors.

\subsection{Scaling Approximation}

After analyzing DA and LMA in terms of pair correlation functions, it
becomes evident that {\it excluded volume} effects are not taken into
account correctly by these approximations, since the exact hard core
conditions, $g_{\alpha\beta}\left(r\right)=0$ for $r<\sigma_{\alpha
\beta}$, are not satisfied. To avoid this defect and obtain reasonably
accurate RDFs of HS mixtures from pure fluid ones with a limited effort, we
propose a {\it scaling approximation}, derived from Eq. (\ref{cs14}) with
the choice

\begin{equation}
\lambda_{\alpha\beta} =\sigma_\mono/\sigma_{\alpha\beta} \qquad \text{and\qquad} \sigma_\mono=\left\langle \sigma^3\right\rangle^{1/3}.
\label{sa1}
\end{equation}

\noindent
Since $\lambda_{\alpha\beta} r_\mono^*=r_{\alpha\beta}^*$, with the
definition $r_{\alpha\beta}^*\equiv r/\sigma_{\alpha\beta}$, our SA for HS
can be written as

\begin{equation}
g_{\alpha\beta}\left(r;\rho,{\bf x};\{\sigma_{\gamma\delta}\}\right)
 \simeq \widehat{g}_\mono\left(r_{\alpha\beta}^*;\eta
\right). \label{sa3}
\end{equation}

\noindent
Note that $\sigma_\mono$ is the same as in DA, while the choice for
$\lambda_{\alpha\beta}$ ensures that, when $r<\sigma_{\alpha\beta}$, one
gets $r_{\alpha\beta}^*<\sigma_\mono$ and, consequently,
$g_{\alpha\beta}(r)=0$. Since excluded volume effects are very important for
the structure of condensed fluids, it is therefore reasonable to expect that
SA is better than both DA and LMA, although it incorrectly assumes that all RDF
values at contact are equal: $g_{\alpha\beta}(\sigma_{\alpha\beta})\simeq
g_\mono(\sigma_\mono)$. Once again the choice for $\sigma_\mono$ ensures
that the RDF of the pure fluid is evaluated at the same packing fraction of
the mixture. From Eqs. (\ref{sm}), (\ref{cs17}) and (\ref{sa1}) one then finds

\begin{equation}
S_M(q)^{SA} = 1 + \sum_{\alpha=1}^p \sum_{\beta=1}^p
  x_\alpha x_\beta ~ w_\alpha(q) w_\beta(q) ~
\frac{\sigma_{\alpha\beta}^3}{\left\langle\sigma^3\right\rangle} ~ \left[ \widehat{S}_\mono(q_{\alpha\beta}^*;\eta
)-1\right], \label{sa4}
\end{equation}

\noindent
with the definition $q_{\alpha\beta}^*\equiv q\sigma_{\alpha\beta}$.
$S_{NN}(q)^{SA}$ is again obtained from the expression for $S_M(q)^{SA}$ by
putting all $w_\nu(q)=1$.

It should be noted that our SA closely resembles, albeit it is not identical
to, the so-called {\it van der Waals one-fluid} (vdW1) {\it
approximation}.\cite{Smith1973,Mcdonald1973,Lee1988} In vdW1 thermodynamics
calculations the mixture is replaced by a single pure fluid, with averaged
potential parameters $\sigma_x,\varepsilon_{x}$ given by van der Waals rules:

\begin{equation}
\left\{
\begin{array}{l}
\sigma_x^3 =
  \sum_\alpha \sum_\beta x_\alpha x_\beta ~ \sigma_{\alpha\beta}^3, \\
\varepsilon_x =
  \sigma_x^{-3}\sum_\alpha \sum_\beta x_\alpha x_\beta ~
    \varepsilon_{\alpha\beta} ~ \sigma_{\alpha\beta}^3.
\end{array}
\right. \label{sa7}
\end{equation}
(of course, there is no $\varepsilon_{x}$ and no dependence on $T$ in the
RDFs of HS fluids). Although often presented in the literature with a
different notation, vdW1 may be regarded as a scaling approximation,
obtainable from Eq. (\ref{cs14}) with $\lambda_{\alpha\beta}
=\sigma_\mono/\sigma_{\alpha\beta}$ and $\sigma_\mono=\sigma_x$. This
choice for $\sigma_\mono$ implies that the vdW1 RDF of the pure fluid is
evaluated at a packing fraction, $\eta_x\equiv (\pi/6)\rho \sigma_x^3$,
which differs from that of the mixture, $\eta \equiv (\pi/6)\rho
\left\langle \sigma^3\right\rangle$. As a consequence, for HS we can write

\begin{equation}
S_M(q)^{vdW1}=1+\sum_{\alpha=1}^p\sum_{\beta=1}^p
x_\alpha x_\beta ~ w_\alpha(q)w_\beta(q) ~
\frac{\sigma_{\alpha\beta}^3}{\sigma_x^3} ~
\left[\widehat{S}_\mono(q_{\alpha\beta}^*;\eta_x)-1\right] .
\label{sa8}
\end{equation}

\subsection{Exact PY solution}

The closed analytical expression for $S_M(q)$ of polydisperse HS in the PY
approximation, $S_M(q)^{PY}$ can be found in Vrij's original paper \cite{Vrij1978} or in Ref. \onlinecite{Gazzillo1997} (with all charges set to
zero). The corresponding expression for $S_{NN}(q)^{PY}$ is simply obtained
by putting $F_\nu(q)=1$ for all form factors in the $S_M(q)^{PY}$ formula.

\subsection{Numerical results}

We tested the results for $S_M(q)$ obtained from DA, LMA, SA and vdW1
against Vrij's $S_M(q)^{PY}$, which is exact within the PY approximation.
Such a comparison requires the evaluation of
$\widehat{S}_\mono(q_\mono^*,\eta_\mono)^{PY}$, for which a simple
analytical expression is available.\cite{March1976}

Using the Schulz distribution to represent the size polydispersity, the
packing fraction of the HS mixture may be written as

\begin{equation}
\eta \equiv (\pi/6)\rho \left\langle \sigma^3\right\rangle =
  (\pi/6)\rho^*\left(1+s^2\right) \left(1+2s^2\right),
\end{equation}

\noindent
where we have chosen $\left\langle\sigma\right\rangle$ as the unit of length,
and defined the dimensionless density
$\rho^*\equiv\rho\left\langle\sigma\right\rangle^3$.

\noindent
Since working with dimensionless variables is very convenient, we also define
$r^* \equiv r/\left\langle\sigma\right\rangle$,
$\sigma^* \equiv \sigma/\left\langle\sigma\right\rangle$ and
$q^* \equiv q\left\langle\sigma\right\rangle$.

The effect of polydispersity on structure factors may be studied by varying
$s$ with either $\eta$ or $\rho^*$ constant. Qualitatively, the results at a
fixed total density and those at a fixed packing fraction are similar. In
Figures \ref{f:Fig1} and \ref{f:Fig2} we present some results for $S_M(q)$ at
fixed packing fraction $\eta=0.3$, with varying polydispersity, $s=0.1$,
$0.3$ and $0.5$ (most of the experimental $s$ values lie in the range $0\div
0.3$). The three distributions of diameters were discretized with a grid size
$\Delta\sigma^*=0.02$, and truncated at $\sigma_\cut^*=1.68$, $3.48$ and
$5.90$ (where $f(\sigma)\Delta \sigma \approx 10^{-8}$). These values of
$\sigma_\cut^*$ correspond to polydisperse mixtures with a number of
components $p=85$, $175$ and $296$, practically intractable with the
available algorithms for solving IEs numerically.

At a fixed $\eta$, calculating $S_M(q)$ or $S_{NN}(q)$ in the DA, LMA and SA
approximations requires the knowledge of
$\widehat{S}_\mono(\widehat{q},\eta)^{PY}$ at
$\widehat{q}=q\left\langle\sigma^3\right\rangle^{1/3}$, $q\sigma_\alpha$ and
$q\sigma_{\alpha\beta}$. In other words, for each $q$ value, a single
evaluation of $\widehat{S}_\mono$ is needed for DA, a number $p$ of
evaluations is required for LMA, and $p(p+1)/2$ evaluations for SA and
wdW1. For wdW1, $\widehat{S}_\mono$ is computed at $\eta_\mono=\eta_x$, rather
than $\eta_\mono=\eta$. To save computer time in LMA, SA and vdW1
calculations, we avoid the repeated evaluations for each $q$ by taking
advantage of the fact that $\widehat{S}_\mono$ does not depend on $q$ and
$\sigma$ separately, but only on their product
$\widehat{q}=q\sigma_{\alpha\beta}=q^*\sigma_{\alpha\beta}^*$. Thus we can
choose a suitable grid size $\Delta q^*$ and a number of points ${\cal N}$, and
calculate $\widehat{S}_\mono(\widehat{q}_i,\eta)^{PY}$ at the grid points
$\widehat{q}_i=i\Delta q^*$ ($i=0,\ldots,{\cal N}-1$) {\it only once}, storing
all values in an array. Of course, the grid points $\widehat{q}_i$ do not
exhaust all the required $q\sigma_{\alpha\beta}$ values. Nevertheless, if the
grid size is small enough, the value of the continuous function
$\widehat{S}_\mono$ at
$\widehat{q}=q^*\sigma_{\alpha\beta}^*$ can be
approximated with that at the nearest grid point, whose index in the array is
simply determined from the ratio $q^*\sigma_{\alpha\beta}^*/\Delta q^*$.
In the worst case, $s=0.5$, to get $S_M(q)$ in a range $0\leq q^*\leq
q_{\max}^*\approx 20$ (a reasonable choice), $\widehat{S}_\mono(\widehat{q})$
must be evaluated in a range $0\leq \widehat{q}\leq
\max(q^*\sigma_\cut^*) \approx 120$. To satisfy such a condition and
get a good grid size in $q^*$-space, $\Delta q^*=q_{\max}^*/{\cal N}$, we
choose $q_{\max}^*=\pi/\Delta \sigma^*=50\pi$ and ${\cal N}=4096$.

Figure \ref{f:Fig1} shows DA and LMA results for $S_M(q)$ versus the exact PY
ones. Similar plots are reported by Pedersen.\cite{Pedersen1994} With
increasing polydispersity, $S_M(q)$ increases in the low-$q$ region, its
first peak is reduced and shifted to smaller $q$ values, and the subsequent
oscillations are progressively washed out. The failure of DA, even at low
polydispersity, is evident not only at small $q$, but also in the first peak
region. LMA is significantly better at small scattering vectors, but does not
reproduce the shape of the first peak correctly.

The corresponding SA and vdW1 results are plotted in Figure \ref{f:Fig2}. Now
the agreement with the exact PY data is surprisingly good in both cases: the
position of the first peak is well reproduced, and its height is only
slightly underestimated. However, SA is globally superior: for $s=$ $0.3$ and
$0.5$ its worst discrepancies are found near the origin, at $q^*\lesssim 2$,
whereas the vdW1 curves are somewhat shifted with respect to the exact ones
on the left side of the first peak. It should be recalled that, unlike vdW1,
SA is evaluated at the true packing fraction of the mixture.

Finally, it is worth looking at the number-number structure factor
$S_{NN}(q)$. In Figure \ref{f:Fig3} a comparison is made among
$S_{NN}(q)^{PY}$, $S_M(q)^{PY}$, and $S_\mono(q)^{PY}$. The results for
$S_{NN}(q)^{PY}$ were obtained from our closed analytical expression, which
is exact within the PY approximation (Griffith {\it et al.} \cite{Griffith1986}
presented similar data, calculated without using a closed
formula \cite{Blum1979} ). For $s=0$, all these structure factors coincide, but
with increasing $s$ the differences become larger and larger. In particular,
$S_{NN}(q)^{PY}$ exhibits a more rapid flattening of the first peak than
$S_M(q)^{PY}$, while its increase in the low-$q$ region is much more dramatic
(see the extreme case $s=0.9$, also included). Since that $S_{NN}(0)$
measures the fluctuations in the total particle density (irrespective of the
species), this behavior of $S_{NN}(q)$ provides more physical information
than that provided by $S_M(q)$. Figure \ref{f:Fig4} then shows the
$S_{NN}(q)$ predicted by SA, compared to $S_{NN}(q)^{PY}$. The performance of
SA is good in the first peak region and beyond, but the approximation fails
in the low-$q$ region and is unable to reproduce the number density
fluctuations correctly. This discrepancy is, however, less important in
$S_M(q)$, which is the structure factor more directly comparable with small
angle scattering data. In fact, multiplying $S_{\alpha\beta}(q)$ by the
product of form factors $w_\alpha(q)w_\beta(q)$ reduces the mentioned defect
(as already seen in Figure \ref{f:Fig2}); moreover, a significant part of the
region near $q=0$ is experimentally unaccessible.

\section{Lennard Jones potential}

From $\varepsilon_\alpha =\varepsilon_{\left\langle \sigma \right\rangle
}\left(\sigma_\alpha /\left\langle \sigma \right\rangle \right)^z$ and
the Berthelot rule, Eqs. (\ref{pcl13}) and (\ref{pl2}), it follows that

\begin{equation}
\varepsilon_{\alpha\beta}^*\equiv \frac{\varepsilon_{\alpha\beta}}{k_BT}=\frac 1{T^*}\left(\frac{\sqrt{\sigma_\alpha \sigma_\beta}}{\left\langle \sigma \right\rangle} \right)^z, \label{lj1}
\end{equation}

\noindent
with $T^*\equiv k_BT/\varepsilon_{\left\langle\sigma\right\rangle}$.

As for HS mixtures, in addition to a density $\rho^*\equiv \rho
\left\langle \sigma \right\rangle^3$, it is convenient to define a second
dimensionless variable $\phi\equiv\rho\left\langle\sigma^3\right\rangle$,
or $\eta_{LJ}\equiv\phi\pi/6$, which plays the same role as the HS
packing fraction $\eta$ (although, rigorously, a LJ particle has no
definite boundary and volume). For the Schulz size distribution,
$\phi=\rho^*\left(1+s^2\right)\left(1+2s^2\right)$.

We performed IE and MD calculations for LJ mixtures at fixed $T^*=1$,
$\phi=0.8$ ($\eta_{LJ}\simeq 0.42$) with $z=2$ and polydispersity parameter
$s=0$ (monodisperse case), $0.1$ and $0.3$. For $s=0$, one has $\phi=\rho^*$,
and the corresponding thermodynamic state, $\rho^*=0.8$, $T^*=1$, is near the
triple point in the LJ phase diagram. \cite{Gazzillo1993} With increasing $s$ at
fixed $\phi$, $\rho^*$ decreases. Taking into account the superiority of SA
with respect to DA, LMA and vdW1 approximation in the polydisperse HS case,
our IE calculations for the LJ potential were restricted only to SA (some
information about the performance of vdW1 for LJ potentials can be found in
works by Hoheisel {\it et al.},\cite{Hoheisel1983,Hoheisel1984} but only for
binary mixtures).

Kofke and Glandt \cite{Kofke1986,Kofke1987} presented similar Monte Carlo results
for a polydisperse LJ model of n-paraffins. Unfortunately, since these
authors started from an activity distribution, rather than a composition
distribution, their simulations are not directly comparable with our IE
calculations.

\subsection{Molecular Dynamics simulations}

The MD simulations employed 1000 particles, in a cube with periodic boundary
conditions, and, using Andersen's stochastic collision
method,\cite{Andersen1980} described a fluid constrained at constant $\phi=0.8$
and maintained in contact with an heat bath. The equations of motion were
integrated using the velocity Verlet algorithm.\cite{Verlet1967,Fox1984} The
atomic diameters $\sigma$ where drawn from a Schulz distribution at the
desired $s$, with well depths given by Eq. (\ref{lj1}). To reduce the sampling
noise, the distribution was periodically regenerated during the simulation. To
avoid truncation effects in the Fourier transforms, $S_M(q)$ was evaluated in
the simulation using directly the definition

\begin{equation}
S_M(q)=\frac{\left\langle \left| \sum_iF_i(q)
  \exp(i{\bf q\cdot r}_i)\right|^2\right\rangle} {\sum_iF_i^2(q)},
\end{equation}

\noindent
where ${\bf q}$ is the wavevector, $q=|{\bf q}|$, ${\bf r}_i$ is the position
of the $i-$th particle and $F_i(q)$ its form factor, Eq. (\ref{scatt}). A
similar expression was used to obtain $S_{NN}(q)$, and we also calculated its
$r$-space counterpart, the number-number correlation function \cite{March1976}
$g_{NN}(r)$.

\subsection{Scaling Approximation and numerical results}

The scaling approximation for the LJ potential is derived from
Eq. (\ref{cs14}) with the choice
$\lambda_{\alpha\beta}=\sigma_\mono/\sigma_{\alpha\beta}$ and
$\sigma_\mono=\left\langle \sigma^3\right\rangle^{1/3}$
(analogous to Eq. (\ref{sa1}) for the HS potential), and

\begin{equation}
\varepsilon_\mono=\varepsilon_{\left\langle \sigma \right\rangle}
\left(\frac{\sigma_\mono} {\left\langle \sigma \right\rangle} \right)^z,
 \label{salj2}
\end{equation}

\noindent
which is consistent with Eqs. (\ref{pcl13}) and (\ref{lj1}). The SA
for LJ therefore reads

\begin{equation}
g_{\alpha\beta}\left(r;\rho,{\bf x},T;\{\sigma_{\gamma\delta}\},
\{\varepsilon_{\gamma\delta} \}\right) \simeq
\widehat{g}_\mono\left(r_{\alpha\beta}^*;\eta_{LJ},T_\mono^*\right).
 \label{sagr}
\end{equation}

\noindent
Note that $T_\mono^*$ decreases with increasing $s$: with Eqs.
(\ref{cs15}), (\ref{salj2}) and the Schulz distribution,
$T_\mono^*=T^*/\left[ (1+s^2)(1+2s^2)\right]^{z/3}$.

To evaluate $\widehat{S}_\mono(q_{\alpha\beta}^*;\eta_{LJ},T_\mono^*)$, we
solved numerically, for each $T_\mono^*$ value, the MHNC-ML integral equation
with ${\cal N}=4096$ and $\Delta r^*=0.02$ (which corresponds to
$q_{\max}^*=\pi/\Delta r^*=50\pi$, just as in the HS case). From the
resulting $\widehat{S}_\mono$, the polydisperse structure factors
$S_M(q)^{SA}$ and $S_{NN}(q)^{SA}$ were finally evaluated from the obvious
analogue of Eq. (\ref{sa4}), in the same manner presented in detail for the
HS potential.

In Figures \ref{f:Fig5} and \ref{f:Fig6} the $S_M(q)^{SA}$ and
$S_{NN}(q)^{SA}$ results are tested against the corresponding MD data. The
overall agreement is good. For $s=0.1$ the positions of maxima and minima and
the shape of both structure factors are correctly predicted; only the height
of the first peak is slightly underestimated. For $s=0.3$ the first peak is
well reproduced in $S_{NN}(q)^{SA}$, but not equally well in
$S_M(q)^{SA}$. On the contrary, $S_{NN}(q)^{SA}$ is completely off in the
region near the origin, where the discrepancy in $S_M(q)^{SA}$ is less
dramatic. Finally, Figure \ref{f:Fig7} shows the behavior of $g_{NN}(r)$:
also in $r-$space SA works quite well. In the IE calculations we employed the
expression \cite{March1976}

\begin{equation}
g_{NN}(r)=\sum_{\alpha=1}^p\sum_{\beta=1}^p x_\alpha x_\beta ~
g_{\alpha\beta}(r).
\end{equation}

\noindent
The MD data for the monodisperse case, $s=0$, have been found to be in
perfect agreement with the Monte Carlo results for $g(r)$ published by
Llano-Restrepo and Chapman \cite{Llanorestrepo1992} (not reported in Figure
\ref{f:Fig7}).

The slightly worse performance of SA in the LJ case with $s=0.3$, with
respect to the HS one with the same polydispersity, depends only on the
higher packing fraction ($\eta_{LJ}=0.42$, whereas $\eta=0.3$ for hard
spheres).

Finally, it is worth mentioning that we also performed MD simulations with
different $z$ values ($z=1$ and 3), and tested variants of the SA with
alternative choices for $\varepsilon_\mono$ (for instance,
$\varepsilon_\mono=\varepsilon_{\left\langle\sigma\right\rangle}$, or
$\varepsilon_\mono=\varepsilon_{\left\langle\sigma\right\rangle}
\left(\sigma_\mono/\left\langle\sigma\right\rangle\right)^3)$. Only very
small changes of peak heights were found in all cases. This relative
insensitiveness to the well depths may perhaps be explained by the well
known fact that at high packing fractions the structure depends essentially
on the repulsive part of potentials, while the attractive forces sensibly
affect only thermodynamics.

An improved SA could be obtained by replacing $T_{{\rm mono}}^{*}$ in
Eq. (\ref{sagr}) with
$T_{\alpha \beta}^{*}\equiv k_BT/\varepsilon_{\alpha\beta}$.
This choice would be akin to the ``mean
density approximation'',\cite{Mansoori1972} used for binary LJ
mixtures.\cite{Hoheisel1983,Hoheisel1984,Shukla1986} However, we have not
attempted its application to polydisperse fluids, since it would require the
evaluation of $\widehat{g}_{{\rm mono}}$ at many  different
reduced temperatures, which is inconvenient when IEs have to be solved
numerically. In the particular case of the LJ potential one could take
advantage of an explicit analytical parametrization for the RDF proposed by
Goldman.\cite{Goldman1979} This extension goes beyond the
scope of the present paper and might be the subject of future work.

\section{Conclusions}

In this paper we have shown that the structure of simple polydisperse fluids
can be successfully predicted by integral equation methods, even when no
analytical solution is available and a numerical one is unpracticable
because of the very large number of components. We have obtained rather
accurate structure factors by solving IEs for appropriate pure fluids and
applying a recipe based upon a corresponding states approach, which assumes
conformality of all RDFs in the polydisperse mixture. Thus the hypothesis that
all RDFs have essentially the same shape, but are scaled with respect to
each other, turns out to be physically sound.

The good performance of the scaling approximation for rather concentrated
systems is due to a correct treatment of excluded volumes. This feature
appears to be lacking in both the decoupling approximation and the local
monodisperse approximation, when these are reformulated in terms of pair
correlation functions. The importance of evaluating the properties of the
reference pure fluid at exactly the same packing fraction of the mixture is
also to be emphasized. This condition is not satisfied by the very similar
vdW1 approximation. A shortcoming of SA is certainly the equality of all RDFs
at contact, but this error appears to be a higher order effect, at least at
high packing fractions.

In conclusion, the proposed SA approximation offers to small angle scattering
experimentalists a simple and valuable tool to predict structure factors or
to fit data, and has proved superior to both DA and LMA. We have shown that
the SA is applicable to different potentials for nonionic fluids. An
extension of SA to polydisperse ionic mixtures will be presented in a
forthcoming paper.

\acknowledgments
The Italian MURST (Ministero dell'Universit\`a e della Ricerca Scientifica e
Tecnologica), the INFM (Istituto Nazionale di Fisica della Materia) and the
University of Bologna (``Finanziamento speciale alle strutture'') are
gratefully acknowledged for financial support. Two of us (AG) and (FC) thank
Pietro Ballone for enlighting discussions.

\begin{figure}[tbp]
\caption{Structure factor $S_{M}(q)$ of hard spheres, for different degrees
of polydispersity $s$, at fixed packing fraction $\eta= 0.3$ (here, and in
all following figures, curves at different $s$ are shifted upwards to avoid
overlapping). Solid lines: Vrij's exact PY results. Dashed lines: local
mononodisperse approximation, LMA. Dotted lines: decoupling approximation,
DA.}
\label{f:Fig1}
\end{figure}

\begin{figure}[tbp]
\caption{Structure factor $S_{M}(q)$ of hard spheres, for different degrees
of polydispersity $s$, at fixed packing fraction $\eta= 0.3$. Solid lines:
Vrij's exact PY results. Dashed lines: scaling approximation, SA. Dotted
lines: van der Waals one-fluid approximation, vdW1.}
\label{f:Fig2}
\end{figure}

\begin{figure}[tbp]
\caption{Comparison of exact PY results for HS structure factors $S_{M}(q)$
and $S_{NN}(q)$, for different degrees of polydispersity $s$, at fixed
packing fraction $\eta= 0.3$. Solid lines: exact PY results for
$S_{NN}(q)$. Dashed lines: exact PY results for $S_{M}(q)$. Dotted lines:
monodisperse structure factor.}
\label{f:Fig3}
\end{figure}

\begin{figure}[tbp]
\caption{Structure factor $S_{NN}(q)$ of hard spheres, for different degrees
of polydispersity $s$, at fixed packing fraction $\eta=0.3$. Solid lines:
exact PY results. Dashed lines: scaling approximation, SA.}
\label{f:Fig4}
\end{figure}

\begin{figure}[tbp]
\caption{Structure factor $S_{M}(q)$ of Lennard-Jones fluids, for different
degrees of polydispersity $s$, at fixed $\phi=0.8$ and $T^*=1$. Points: molecular dynamics data. Lines: scaling approximation, SA.}
\label{f:Fig5}
\end{figure}

\begin{figure}[tbp]
\caption{Structure factor $S_{NN}(q)$ of Lennard-Jones fluids, for different
degrees of polydispersity $s$, at fixed $\phi=0.8$ and $T^*=1 $. Points: molecular dynamics data. Lines: scaling approximation, SA.}
\label{f:Fig6}
\end{figure}

\begin{figure}[tbp]
\caption{Number-number correlation functions $g_{NN}(r)$ of Lennard-Jones
fluids, for different degrees of polydispersity $s$, at fixed $\phi=0.8$
and $T^*=1$. Points: molecular dynamics data. Lines:
scaling approximation, SA.}
\label{f:Fig7}
\end{figure}



\begin{thebibliography}{10}

\bibitem{Pusey1991} 
P. N. Pusey, 
in {\it Liquids Freezing and Glass Transition: II, Les Houches Sessions 1989}, 
eds. J. P. Hansen, D. Levesque and J. Zinn-Justin (North-Holland, Amsterdam, 1991) pp. 763-942.

\bibitem{Nagele1996} 
G. N\"{a}gele, 
Phys. Rep. {\bf 272}, 215 (1996).

\bibitem{Gazzillo1994} 
D. Gazzillo, 
Molec. Phys. {\bf 83}, 1171 (1994).

\bibitem{Gazzillo1995} 
D. Gazzillo, 
Molec. Phys. {\bf 84}, 303 (1995).

\bibitem{Vrij1978} 
A. Vrij, 
J. Chem. Phys. {\bf 69}, 1742 (1978); {\bf 71}, 3267 (1979).

\bibitem{Gazzillo1997} 
D. Gazzillo, A. Giacometti and F. Carsughi, 
J. Chem. Phys. {\bf 107}, 10141 (1997).

\bibitem{Aguanno1991} 
B. D' Aguanno and R. Klein, 
J. Chem. Soc. Faraday Trans. {\bf 87}, 379 (1991).

\bibitem{Aguanno1992} 
B. D' Aguanno and R. Klein, 
Phys. Rev. A {\bf 46}, 7652 (1992).

\bibitem{Smith1973} 
W. R. Smith, 
in {\it Statistical Mechanics, A Specialist Periodical Report, Vol. 1} 
(The Chemical Society, London, 1973) Chapter 2.

\bibitem{Mcdonald1973} 
I. R. McDonald, 
in {\it Statistical Mechanics, A Specialist Periodical Report, Vol. 1} 
(The Chemical Society, London, 1973) Chapter 3.

\bibitem{Lee1988} 
L. L. Lee, 
{\it Molecular Thermodynamics of Nonideal Fluids} 
(Butterworths, Boston, 1988).

\bibitem{Kotlarchyk1983} 
M. Kotlarchyk and S. -H. Chen, 
J. Chem. Phys. {\bf 79}, 2461 (1983).

\bibitem{Pedersen1994} 
J. S. Pedersen, 
J. Appl. Cryst. {\bf 27}, 595 (1994).

\bibitem{Hansen1986} 
J. P. Hansen and I. R. Mc Donald, 
{\it The Theory of Simple Liquids} 
(Academic, London, 1986).

\bibitem{Gazzillo1993} 
D. Gazzillo and R. G. Della Valle, 
J. Chem. Phys. {\bf 99}, 6915 (1993).

\bibitem{Lebowitz1964} 
J. L. Lebowitz, 
Phys. Rev. {\bf A133}, 895 (1964).

\bibitem{Malijevsky1987} 
A. Malijevsky and S. Labik, 
Molec. Phys. {\bf 60}, 663 (1987).

\bibitem{Labik1989} 
S. Labik and A. Malijevsky, 
Molec. Phys. {\bf 67}, 431 (1989).

\bibitem{Rosenfeld1986} 
Y. Rosenfeld and L. Blum, 
J. Chem. Phys. {\bf 85}, 2197 (1986).

\bibitem{Ashcroft1967} 
N. W. Ashcroft and D. C. Langreth, 
Phys. Rev. {\bf 156}, 685 (1967).

\bibitem{Bhatia1970} 
A. B. Bhatia and D. E. Thornton, 
Phys. Rev. {\bf B2}, 3004 (1970).

\bibitem{Kofke1986} 
D. A. Kofke and E. D. Glandt, 
Fluid Phase Equilibria {\bf 29}, 327 (1986).

\bibitem{Kofke1987} 
D. A. Kofke and E. D. Glandt, 
J. Chem. Phys. {\bf 87}, 4881 (1987).

\bibitem{Beurten1981} 
P. van Beurten and A. Vrij, 
J. Chem. Phys. {\bf 74}, 2744 (1981).

\bibitem{Abramowitz1972} 
M. Abramowitz and I. A. Stegun, 
{\it Handbook of Mathematical Functions} 
(Dover, New York, 1972).

\bibitem{Rowley1994} 
R. L. Rowley, 
{\it Statistical Mechanics for Thermophysical Property Calculations} 
(Prentice Hall, New Jersey, 1994).

\bibitem{Lucas1991} 
K. Lucas, 
{\it Applied Statistical Thermodynamics} 
(Springer-Verlag, Berlin, 1991).

\bibitem{Parrinello1974} 
M. Parrinello, M. P. Tosi and N. H. March, 
Proc. Roy. Soc. {\bf A341}, 91 (1974).

\bibitem{Ballone1986} 
P. Ballone, G. Pastore, G. Galli and D. Gazzillo, 
Molec. Phys. {\bf 59}, 275 (1986).

\bibitem{Rogers1984} 
F. J. Rogers and D. A. Young, 
Phys. Rev. {\bf A30}, 999 (1984).

\bibitem{March1976} 
N. H. March and M. P. Tosi, 
{\it Atomic Dynamics in Liquids} 
(MacMillan, London, 1976).

\bibitem{Griffith1986} 
W. L. Griffith, R. Triolo and A. L. Compere, 
Phys. Rev. {\bf A33}, 2197 (1986).

\bibitem{Blum1979} 
L. Blum and G. Stell, 
J. Chem. Phys. {\bf 71}, 42 (1979).

\bibitem{Hoheisel1983} 
C. Hoheisel, U. Deiters and K. Lucas, 
Molec. Phys. {\bf 49}, 159 (1983).

\bibitem{Hoheisel1984} 
C. Hoheisel and K. Lucas, 
Molec. Phys. {\bf 53}, 51 (1984).

\bibitem{Andersen1980} 
H. C. Andersen, 
J. Chem. Phys. {\bf 72}, 2384 (1980).

\bibitem{Verlet1967} 
L. Verlet, 
Phys. Rev. {\bf 159}, 98 (1967).

\bibitem{Fox1984} 
J. F. Fox and H. C. Andersen, 
J. Phys. Chem. {\bf 88}, 4019 (1984).

\bibitem{Llanorestrepo1992} 
M. Llano-Restrepo and W. G. Chapman, 
J. Chem. Phys. {\bf 97}, 2046 (1992).

\bibitem{Mansoori1972} 
G. A. Mansoori and T. W. Leland, 
J. Chem. Soc. Faraday Trans. II {\bf 8}, 320 (1972).

\bibitem{Shukla1986} 
K. P. Shukla, M. Luckas, H. Marquart and K. Lucas, 
Fluid Phase Equil. {\bf 26}, 129 (1986).

\bibitem{Goldman1979} 
S. Goldman, 
J. Phys. Chem. {\bf 83}, 3033 (1979).

\end{thebibliography}
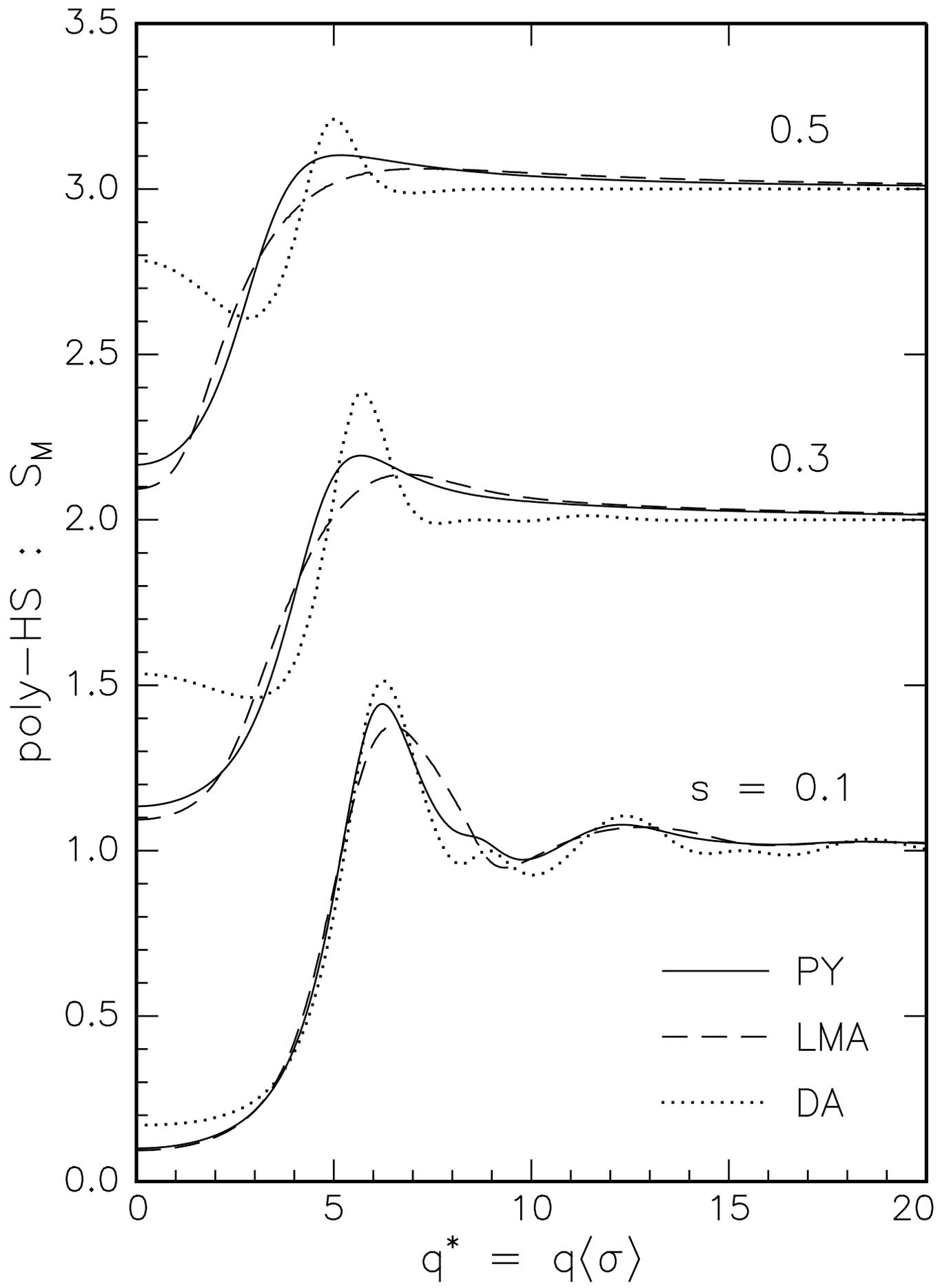
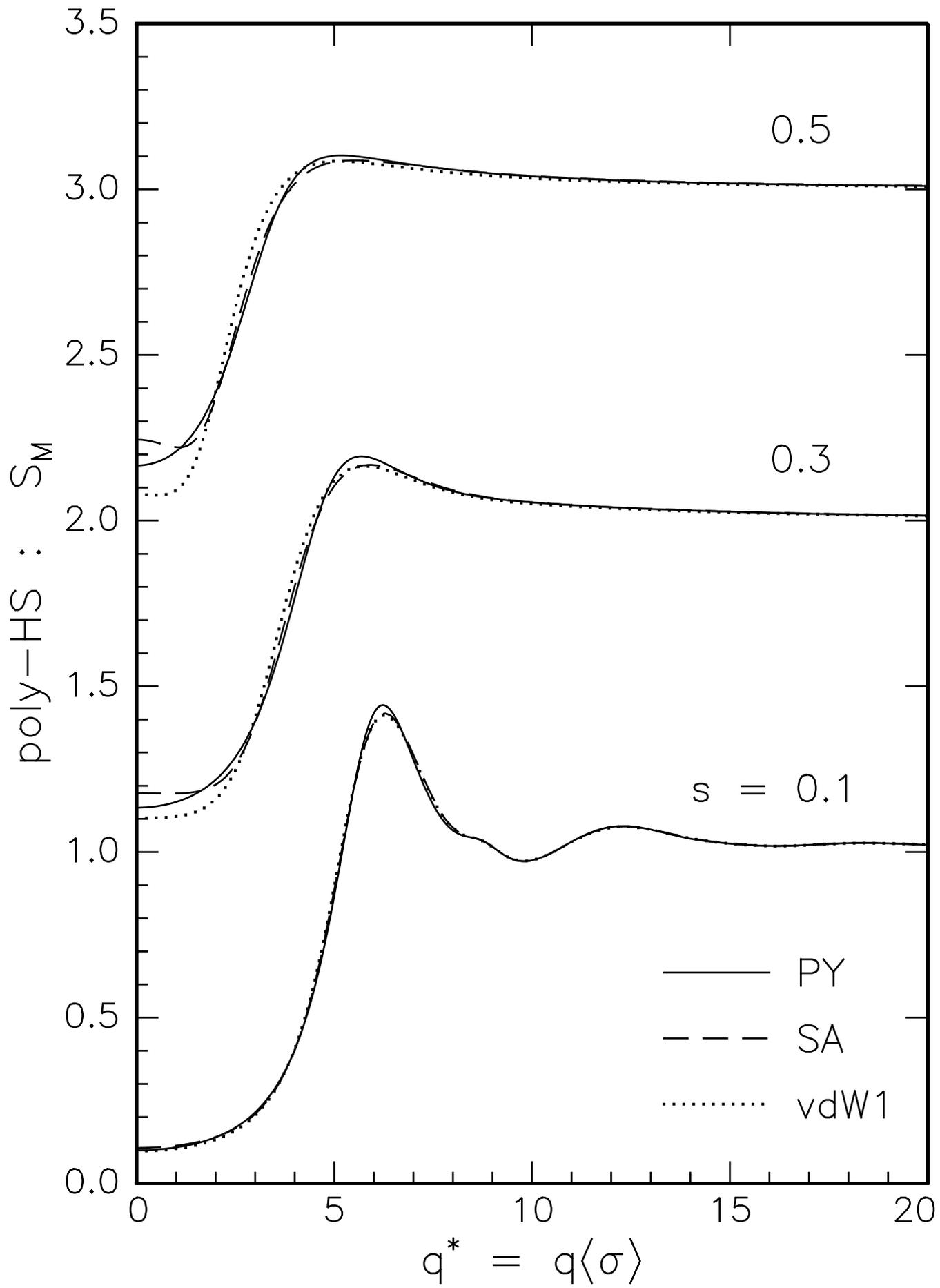
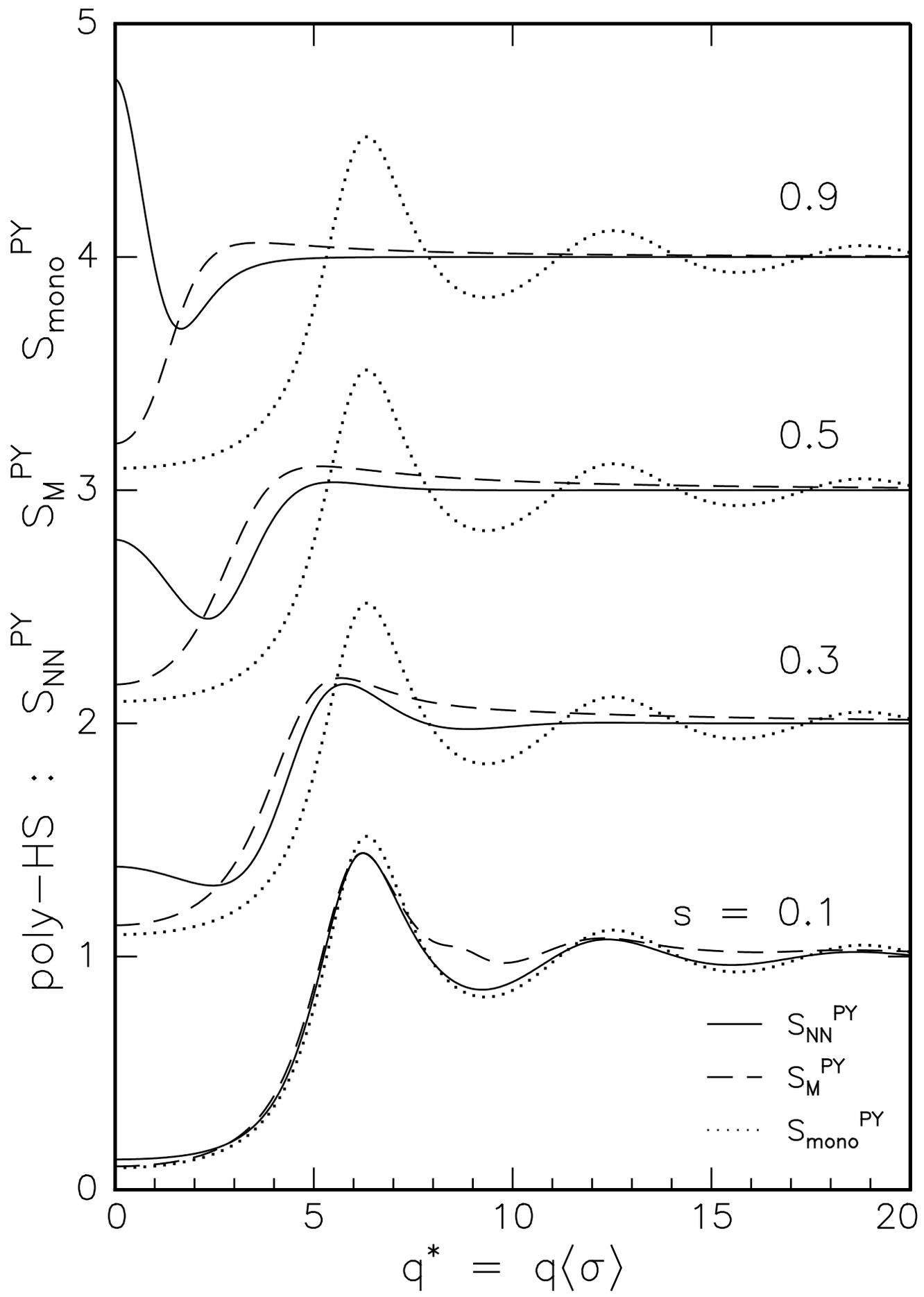
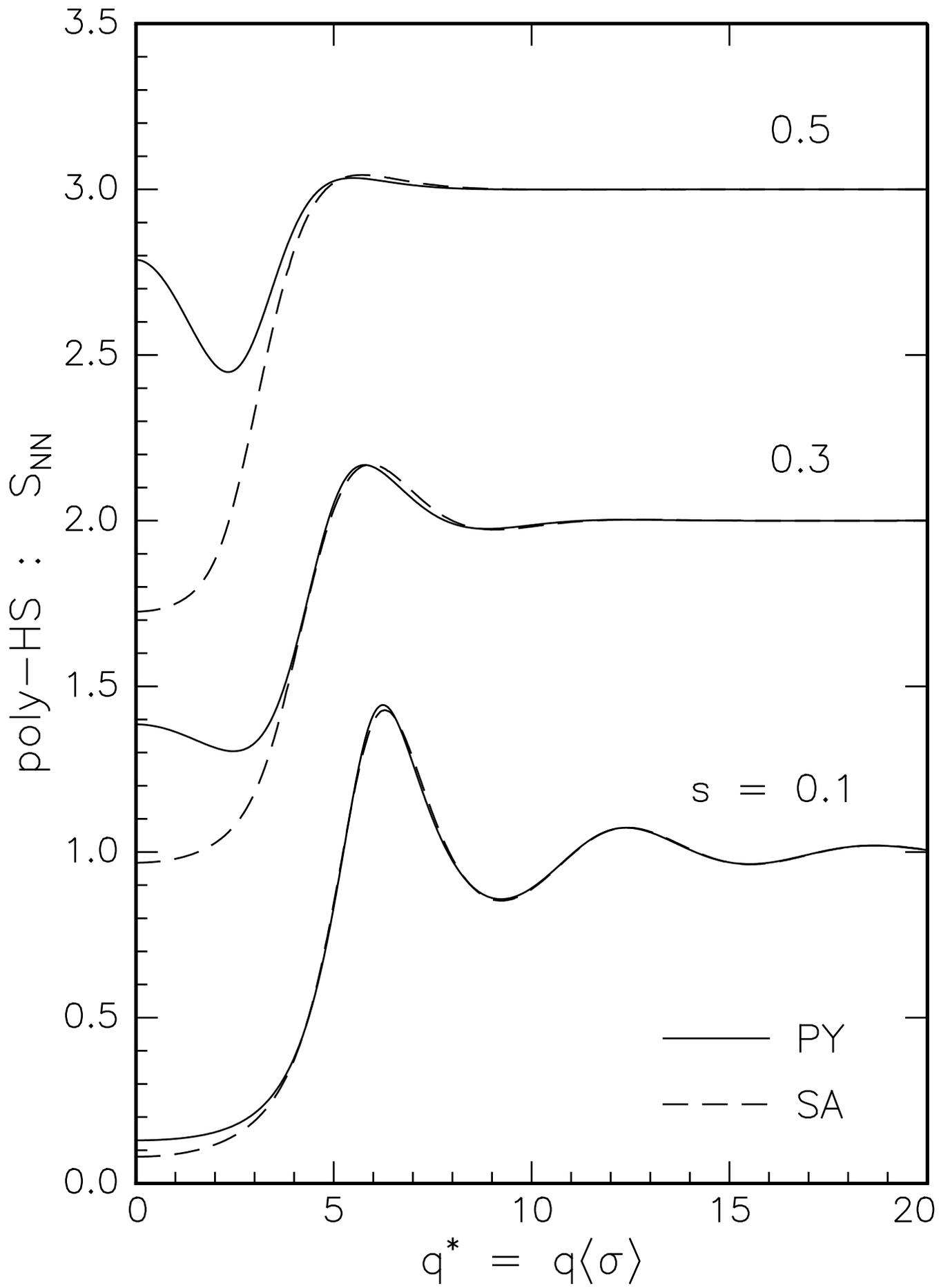
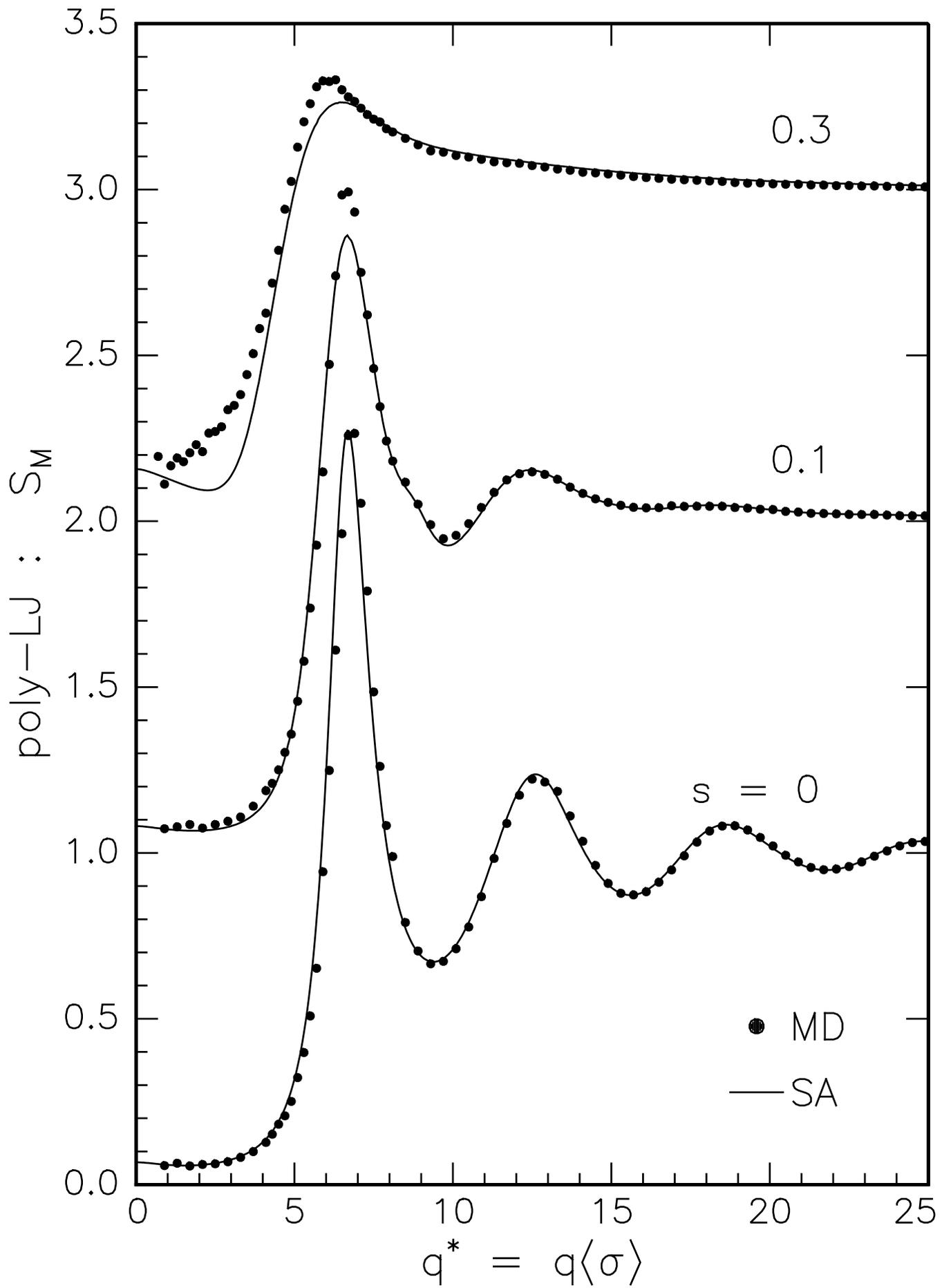
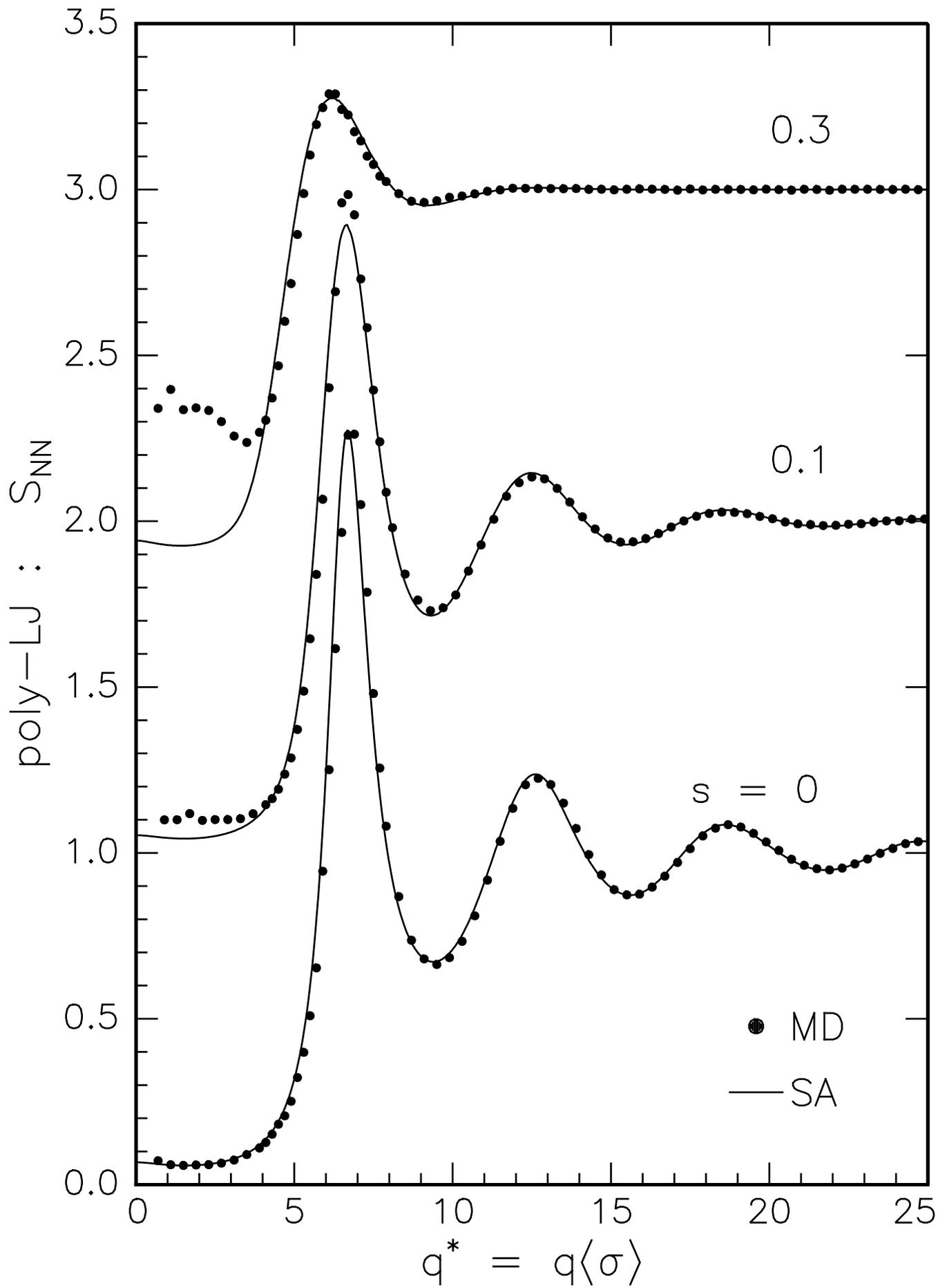
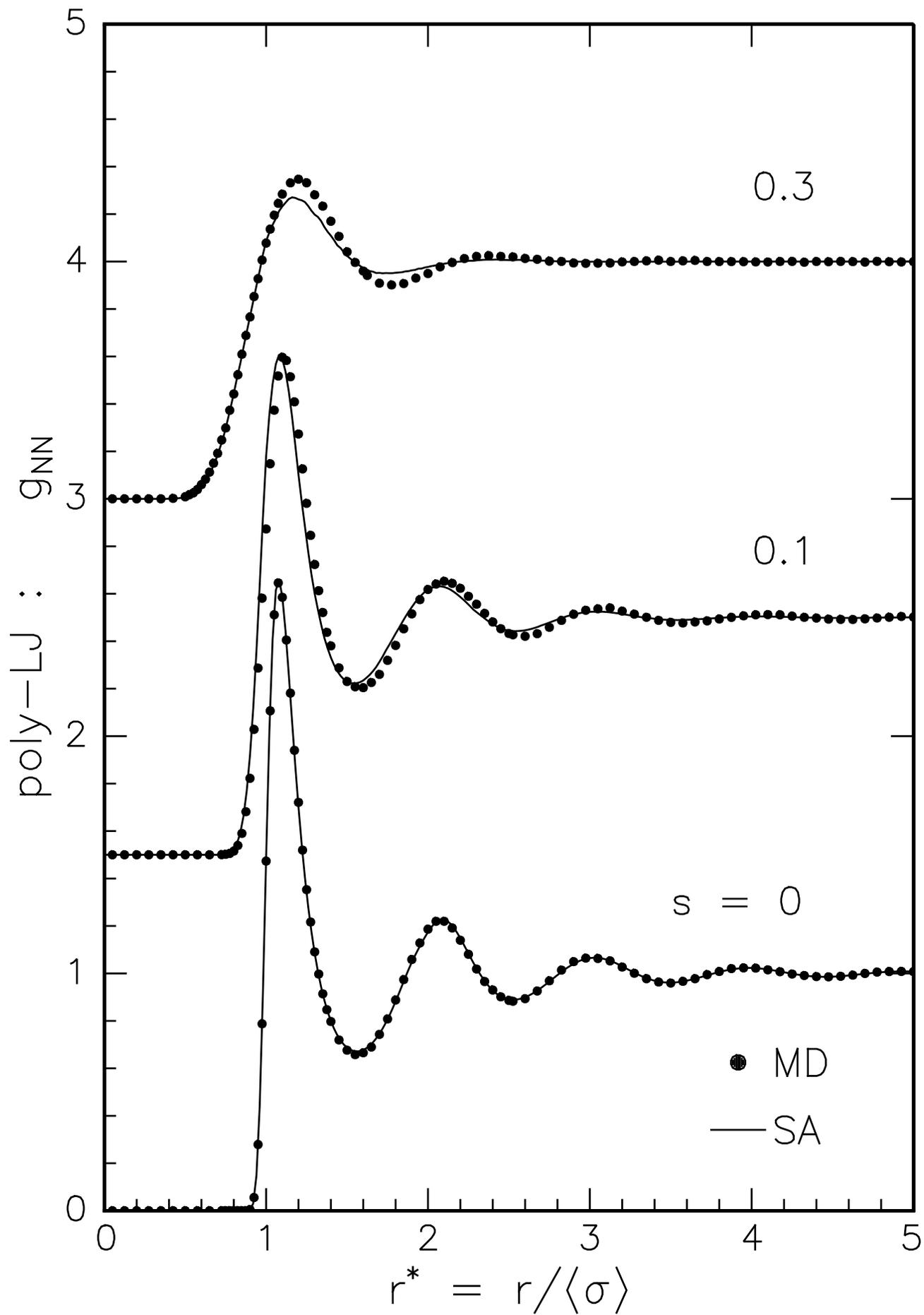
\end{document}